\documentclass[10pt,twocolumn,journal,compsoc]{IEEEtran}
\usepackage{ifpdf}

\ifCLASSOPTIONcompsoc
  \usepackage[nocompress]{cite}
\else
  \usepackage{cite}
\fi

\ifCLASSINFOpdf
  \usepackage[pdftex]{graphicx}
  \DeclareGraphicsExtensions{.pdf,.jpeg,.png}
\else
  \usepackage{graphicx}
  \DeclareGraphicsExtensions{.pdf,.jpeg,.png}
\fi
 
\usepackage[nolist]{acronym}
\usepackage{multirow}
\usepackage{graphicx}
\usepackage{caption}
\usepackage{amsmath}
\usepackage{color}

\usepackage{enumitem}

\usepackage{hyperref}
\usepackage{upgreek}
\usepackage[tight,footnotesize]{subfigure}

\definecolor{orange}{RGB}{255, 165, 0}

\acrodef{ARP}{Access Reservation Protocol}
\acrodef{UE}{User Entitity}
\acrodef{RAO}{Random Access Opportunity}
\acrodef{MTC}{Machine Type Communication}
\acrodef{TTI}{Transmission Time Interval}
\acrodef{LTE}{Long Term Evolution}

\begin{document}
%

\title{Wireless Access for Ultra-Reliable Low-Latency Communication (URLLC): Principles and Building Blocks}

\author{
  \IEEEauthorblockN{%
  Petar~Popovski\IEEEauthorrefmark{1},
  Jimmy~J.~Nielsen\IEEEauthorrefmark{1}, %
  \v Cedomir~Stefanovi\' c\IEEEauthorrefmark{1},
  Elisabeth~de~Carvalho\IEEEauthorrefmark{1}, %
  Erik~Str\"om}\IEEEauthorrefmark{2},
  Kasper~F.~Trillingsgaard\IEEEauthorrefmark{1}, %
  Alexandru-Sabin~Bana\IEEEauthorrefmark{1}, %
  Dong~Min~Kim\IEEEauthorrefmark{1}, %
  Radoslaw~Kotaba\IEEEauthorrefmark{1}, %
  Jihong~Park\IEEEauthorrefmark{1}, %
  Ren\'{e}~B.~S{\o}rensen\IEEEauthorrefmark{1} \\
\vspace{6pt}
\IEEEauthorblockA{\IEEEauthorrefmark{1}Dept.~of Electronic Systems, Aalborg University, 9220 Aalborg, Denmark\\
Emails: %
\{petarp,jjn,cs,edc,kft,asb,dmk,rak,jihong,rbs\}@es.aau.dk, %
} \\
\IEEEauthorrefmark{2}Dept.~of Electrical Engineering, Chalmers
Univ.~of Technology, 412~96 Gothenburg, Sweden\\
Email: erik.strom@chalmers.se 
}
\maketitle

\begin{abstract}
Ultra-reliable low latency communication (URLLC) is an important new feature brought by 5G, with a potential to support a vast set of applications that rely on mission-critical links.
In this article, we first discuss the principles for supporting URLLC from the perspective of the traditional assumptions and models applied in communication/information theory. 
We then discuss how these principles are applied in various elements of the system design, such as use of various diversity sources, design of packets and access protocols.
The important messages are that there is a need to optimize the transmission of signaling information, as well as a need for a lean use of various sources of diversity.
\end{abstract}

\IEEEpeerreviewmaketitle

\section{Introduction}
\label{sec:introduction}

The big difference between 5G and the previous generations of mobile wireless systems is that 5G is natively addressing two generic modes of Machine-Type Communications (MTC): Ultra-Reliable Low-Latency Communication (URLLC) and massive MTC (mMTC).
URLLC is arguably the most innovative feature brought in 5G, as it will be used for mission critical communications, like reliable remote action with robots or coordination among vehicles.
Ultra-reliable communication~\cite{popovski2014ultra} is potentially an enabler of a vast set of applications, some of which are yet unknown.
To put this in perspective, wireless connectivity and embedded processing have significantly transformed many products by expanding functionality and transcending the traditional product boundaries~\cite{porter2014smart}; e.g., a product stays connected to its manufacturer through its lifetime for maintenance and update.
Ultra-reliable connectivity brings this transformation to the next level: Once a system designer can safely assume that wireless connectivity is ``truly anywhere and anytime'', e.g.~guaranteed  $ > 99.999 \, \%$ of the time, the approach to system design and operation changes fundamentally.
An example is Industrie 4.0, where different parts of an object or a machine need not to be physically attached as long as they can use mission-critical ultra-reliable links to work in concert towards accomplishing a production task.

In this paper, we first describe the principles for achieving wireless URLLC, relating them to the traditional assumptions in information and communication theory and elaborating why a new view is required.
We then describe several important building blocks of a wireless communication system for supporting URLLC connections: framing/packetization, use of diversity, network topology and access protocols. The objective of this article is to describe their properties as essential ingredients in practically any URLLC solution, rather than combining them into a full proposal.

\section{Communication-Theoretic Principles for URLLC}
\label{sec:CommTheory}

The simple, but seminal communication model by Shannon~\cite{shannon1948mathematical} captures the essential stochastic nature of a communication system.
The key information-theoretic result is that, given sufficiently long time and sufficiently many communication channel uses, one can obtain almost a deterministic, error-free data transmission whose rate is dictated by the channel capacity.
Here ``sufficiently many'' means that the law of large numbers (LLN) averages out the stochastic variations.
This is challenged in URLLC in at least three aspects:
\begin{enumerate}
\item Due to the latency constraints, the number of available channel uses is limited, such that the LLN cannot be put to work and offer arbitrarily high reliability.
\item Transmission of the actual data is only one ingredient of the whole communication protocol, which involves transmission/exchange of metadata as well as other auxiliary procedures, such as channel estimation, packet detection, additional protocol exchanges, etc.
\item The performance that can be guaranteed depends on the model used during the design and URLLC requires that the models are considered in regimes not treated previously (e.g. very rare events).
\end{enumerate}
In the following we elaborate on the principles to address these aspects which set the basis for the building blocks and the associated research challenges.

\subsection{Latency Constraints}
\label{sec:LC}
  
\begin{figure}
    \centering
    \includegraphics[width=8.3cm]{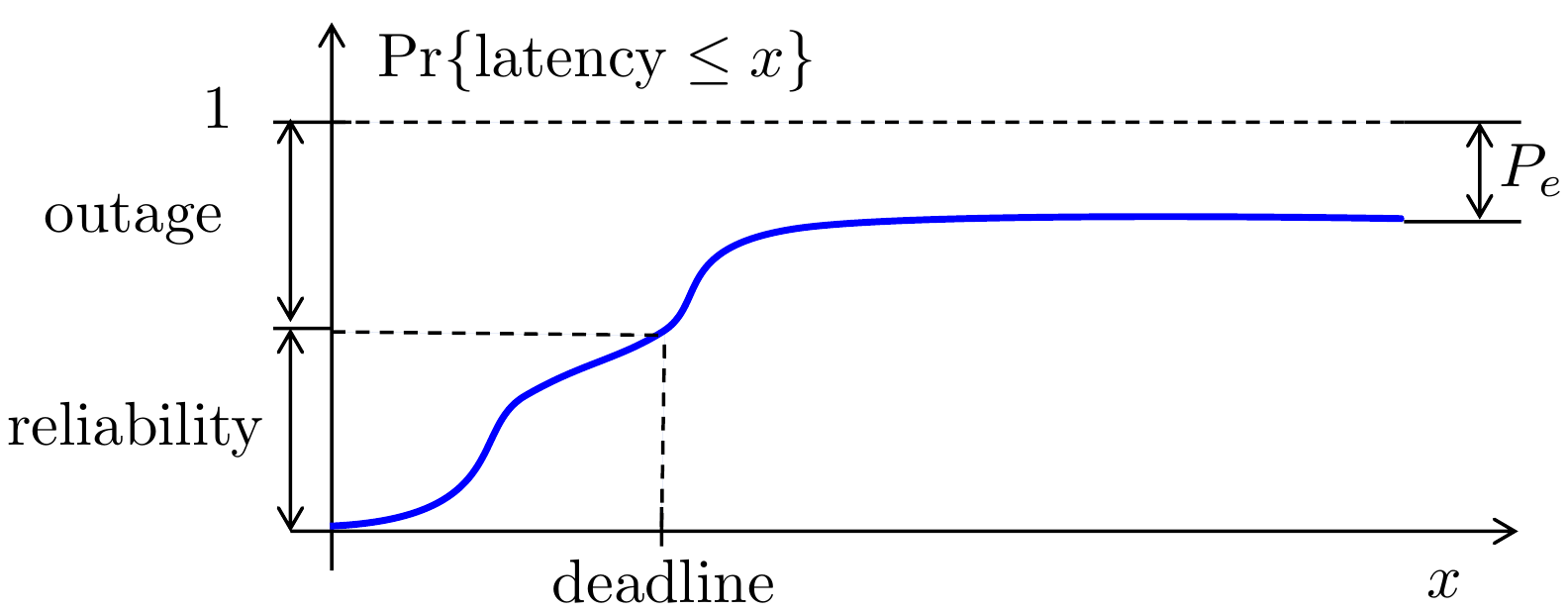}
    \caption{Relation between outage, reliability, latency, and deadline.}
    \label{fig:LatencyReliability}
	\vspace{-0.5cm}
\end{figure}

Latency is defined as the delay a packet (containing a certain number of data bits) experiences from the ingress of a protocol layer at the transmitter to the egress of the same layer at the receiver.
Some packets will be dropped, i.e., never delivered, due to buffer overflows, synchronization failures, etc.
Moreover, we assume that packets that are decoded in error are also dropped---either by the protocol itself or by higher layers. 
Using the convention that dropped packets have infinite latency, we can define the  \emph{reliability} as the probability that the latency does not exceed a pre-described deadline. Fig.~\ref{fig:LatencyReliability} shows the generic requirement in terms of latency and reliability, applicable not only to point-to-point link, but also arbitrary communication setup. The exact numbers on the deadline and the reliability are application dependent.
We note that the latency cumulative distribution function (CDF) asymptote is equal to $1-P_e$, where $P_e$ is the probability of packet drop or packet error. 

Clearly, high reliability implies low $P_e$, but the opposite is not necessarily true, as in URLLC we need to achieve low $P_e$ in a time duration limited by the deadline.
The number of available channel uses is (approximately) proportional to the product of the time duration and the bandwidth of the transmitted signal.
Hence, by increasing the bandwidth, we obtain two advantages: more available channel uses and (typically) more frequency diversity. 
Increasing bandwidth enables us to decrease the channel use time duration, or to keep the duration fixed and to increase the number of channel uses in frequency. 
The trade-offs arising in relation to the definition of time-frequency resources are captured in the flexible numerology used to design the 5G frames \cite{pedersen2015flexible}.  

It is important to note a conceptual difference between increasing the channel uses in time vs.~frequency. 
Assume that Alice sends a packet to Bob by using a common packet structure in which data is preceded by metadata. 
Let the packet transmission consume $N = N_M + N_D$ channel uses, where $N_M$ are intended for metadata and $N_D$ for data. 
If the $N_M$ metadata channel uses precede the data channel uses, then after decoding the metadata, Bob
can decide whether to continue to decode the data from the remaining $N_D$ channel uses (if he is the intended recipient of the data) or to shut down the receiver and save energy. 
Bob cannot save energy in the same way if these $N$ channel uses occur in parallel in frequency, as he needs to receive all symbols before deciding if the packet is
intended for him. 
This follows the intuition that higher reliability necessarily leads to higher energy expenditure.

Besides frequency, URLLC can rely on other types of diversity, such as access point diversity due to densification, spatial diversity due to massive number of antennas and interface diversity.
Further elaboration on these is given in Section~\ref{sec:DiversitySources}.

\subsection{Metadata, Auxiliary Procedures and Protocol Exchanges}
\label{sec:MAPPE}

The capacity results of information theory implicitly assume that when Alice transmits data to Bob, both of them know that the transmission is taking place as well as when it starts and ends.
In practice, this information needs to be conveyed through transmission of metadata (control information).
When the size of the data is much larger than the metadata, as in the classical information-theoretic setup, the amount of resources (channel uses) spent on sending metadata is negligible.
Moreover, it is assumed that the number of channel uses for metadata $N_M$ is sufficiently large to guarantee high reliability, while it still holds that $N_M \ll N_D$.
This does not hold in URLLC, since the data size is often small and comparable to the metadata size, and one explicitly needs to optimize the coding/transmission of metadata.

Further, considering the high reliability levels treated in URLLC, such as e.g. $>99.999\%$, one can no longer assume that the metadata transmission, as well as all auxiliary procedures, are perfectly reliable.
To illustrate this, consider that the probability of success for a given data packet $p$, denoted by $P_S(p)$, is a \emph{product} of the success probabilities for the data $P_{S}(D)$, metadata $P_{S}(M)$, and the auxiliary procedures $P_S(A)$:
\begin{equation}
P_S(p)=P_S(A) P_{S}(M) P_{S}(D)
\end{equation}
This calculation assumes that each procedure is executed independently of the others and, thus each of them designed separately to take place over dedicated communication resources.
However, in principle, one can gather all communication resources and apply a joint design of the three elements: auxiliary (A), metadata (M) and data (D).
Denote the highest probability of success that can be obtained in that case by $Q_S(p, AMD)$.
Clearly, $Q_S(p,AMD) \geq P_S(p)$, since $P_S(p)$ is obtained by using a specific instance in which A, M, and D are separated.
Why then do we not always use joint design of A, M, and D?
This is due to the layered approach to the communication system design, but also due to energy consumption, also discussed in relation to frequency diversity.
Namely, in the common system design, the decoding of data and metadata is causally dependent on the successful completion of the auxiliary procedures (e.g. detection that the packet is there), and, likewise, the decoding of data is causally dependent on the successful decoding of metadata.
If the receiver Bob detects that there is a packet, it proceeds to decode the metadata and, if the packet is relevant, to decode the data.
When A, M and D are not separated by design, then Bob needs to perform all the decoding steps and spend energy, although the received data may not be relevant for him.

The above discussion concerns a single packet transmission from Alice to Bob.
However, the communication protocols often use multiple exchanges between the communicating parties.
One example is user authentication.
As another example, if Bob is a base station (BS) to which Alice wants to transmit, then Bob should send a packet that grants access to Alice, such that Alice can send her data\footnote{This type of coordinated access is used in for example 3GPP cellular systems, but not in typical Wi-Fi deployments.}.
In a simple example, assume that Bob needs to grant access to Alice via packet $p_1$, Alice sends her packet $p_2$ to Bob and finally Bob sends an acknowledgement $p_3$.
The probability of success for the packet $p_i$ is denoted by $P_S(p_i)$, incorporating in it the data and the auxiliary procedures.
Here we cannot do the trick of jointly encoding $p_1, p_2,$ and $p_3$, since each of them is sent by a different party!
The overall probability of success is:
\begin{equation}
P_S=P_S(p_1)P_S(p_2)P_S(p_3)
\end{equation}
such that every additional protocol step decreases the overall reliability.
This has been noticed by researchers, giving rise to grant-free access protocols, see Section~\ref{sec:AccessProtocols}.
This simple analysis also shows that a systematic redesign of the protocols is required when considering the ultra-reliability regime.

The design of packets and access protocols in URLLC regime is further discussed in Sections~\ref{sec:Packetization} and \ref{sec:AccessProtocols}.

\subsection{Use of Appropriate Stochastic Models}

The nature of communication systems and the Shannon-like stochastic models can be used to provide reliability guarantees, provided that the model accurately captures the statistics of all relevant factors.
A communication engineer usually models ``known unknowns''\footnote{Borrowed from the famous quote
  by D. Rumsfeld: ``There are known knowns; there are things we know we know. We also know there are
  known unknowns; that is to say we know there are some things we do not know. But there are also
  unknown unknowns -- the ones we don't know we don't know.''}, but the challenge of URLLC is that
they require modeling of factors occurring very rarely (e.g. with probability of $10^{-6}$) within
the packet duration, if the target reliability is higher (e.g. outage probability in the same period  $<10^{-7}$).
Hence, there is a need to consider factors that so far have been treated as ``unknown unknowns'' in wireless design and performance evaluation.

Specifically, consider a simple model for the signal at a single-antenna receiver:
\begin{equation}
y=hx+z+w
\end{equation}
where $x$ is the transmitted signal, $h$ is the channel coefficient, $z$ is the noise, and $w$ is the interference.
By selecting licensed spectrum, the designer makes $w$ a known unknown.
The noise $z$ is there to represent the stochastic fluctuations, but it is still a known unknown, as its variance is upper-bounded.
If $h$ is known, $w=0$, and the noise is Gaussian, we get the classical Gaussian channel often used to benchmark coding and transmission techniques.
However, accurate knowledge of $h$ or its (tail) statistics is critical for URLLC. 
Using very conservative estimates for the random factors $h$, $z$, and $w$ to be able to guarantee high reliability may lead to very large margins in terms of transmission power or infrastructure.
Therefore, proper stochastic models of the wireless environment are crucial for making URLLC affordable.

\section{Framing and Packetization} 
\label{sec:Packetization}

As discussed above, when the sizes of the preamble, the metadata, and the data are comparable, it is no longer obvious that the conventional frame or packet structure is close to optimal.
In this regime, the channel capacity becomes an inaccurate metric for assessing the necessary blocklength required to achieve a certain reliability.
Instead, an essential quantity is the maximum coding rate for which \cite{Polyanskiy2010b} (and references therein) developed nonasymptotic bounds and approximations of.
For AWGN channels, the key result from these works states that the maximum coding rate is subject to a back-off from the capacity that is approximately proportional to the square root of the blocklength.

A recent study \cite{short_codes_mismatched_csi} has shown that for cases when CSI is unknown and the channel uses are limited, there is an optimal size of the preamble used for CSI acquisition, which depends on the reliability requirement, SNR, frame length and data rate.
This suggests that there might also be an optimal trade-off between the amount of channel uses used for detection and decoding, which may also depend on the reliability requirement, SNR and the available channel uses.
In case the optimal size for the preamble becomes considerably large, joint encoding of training symbols and data symbols could prove to be a more suitable alternative for achieving the latency-reliability requirements. This is in the spirit of joint design of the auxiliary procedures and data/metadata, discussed before. 

\begin{figure}[tb]
  \begin{center}
  \subfigure[]{
  \includegraphics[width=0.45\columnwidth]{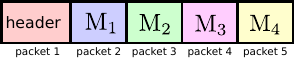}
  }
  \subfigure[]{
  \includegraphics[width=0.45\columnwidth]{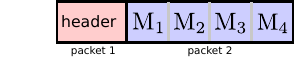}
  }
  \end{center}
  \caption{Two approaches to structuring a frame with a header and messages $\text{M}_1,\ldots,\text{M}_4$ to four devices. (a) Each message is encoded into a separate pacekt. (b) All messages are jointly encoded in a single packet.}
  \label{fig:fbl}
\end{figure}

Furthermore, the insights gained from finite blocklength information theory also allow for rethinking the frame structure in multiuser systems. Here we show how this structure can be changed for downlink transmissions to URLLC devices.
Specifically, consider a wireless system serving multiple URLLC devices with short packets using TDMA.
The BS serves the devices in frames with the aim of delivering independent messages to each device with a certain reliability.
In the conventional approach, depicted in Fig.~\ref{fig:fbl}(a), the BS encodes each message into separate packets and organizes them in a frame with a header containing pointers to each packet.
This approach is optimal from an information-theoretic perspective when the messages are large, because each message can be encoded with a rate close to the channel capacity.
When the messages are small, however, the results from finite blocklength information theory imply that the rate is subject to a back-off from the channel capacity that is inversely proportional to the square-root of the packet length, which is a significant penalty for short packet communications.

As an alternative to the conventional frame structure for downlink broadcast, the transmitter can jointly encode all messages into one packet, thereby leveraging the improved achievable rates when encoding larger messages. As a result, all messages can be delivered with the same reliability, but with a shorter frame.
We depict this approach in Fig.~\ref{fig:fbl}(b).
The approach is not uniformly better than the conventional one, though, as it requires that each device receives and decodes the full packet containing all messages.
From a device perspective, this implies increased power consumption, which is not desirable for devices with power-constraints.
Finally, the two approaches can be considered the extremes of a trade-off between frame duration and power consumption at the devices.
In this context, finite blocklength information theory can help in finding the optimal operating point on this trade-off curve, a problem addressed in greater detail in \cite{Trillingsgaard2017a}.


\section{Types of Diversity}
\label{sec:DiversitySources}
Diversity with respect to paths, can be achieved both through using multiple antennas and by using multiple communication interfaces and technologies. These complimentary techniques are presented in the following.

\subsection{Multi-Antenna Diversity} 

It is well understood that multiple antennas at the BS or terminals are instrumental to guarantee reliable and low latency communications.
The extreme number of spatial degrees of freedom present in massive MIMO can potentially be a significant contributor to URLLC. Indeed, the remarkable properties of massive MIMO that are tailored for URLLC are:
\begin{itemize}
    \item Very high SNR links.
    \item Quasi-deterministic links, quasi-immune to fading.
    \item Extreme spatial multiplexing capability.
\end{itemize}
The first property occurs due to the array gain.
Along with the second property, it relaxes the need for strong coding schemes, hence maintaining high reliability for shorter packets, and can dramatically reduce retransmission occurrences.
The second and third properties are each grounded in the ability of multiple antennas to create spatial diversity paths.
With hundreds of antennas at a massive BS, hundreds of spatial diversity paths can be created, if the propagation channel offers enough scattering.
In practice, if the propagation channel provides an order of tens of diversity paths, it is sufficient to offer statistically stable links.

Nevertheless, the benefits of massive MIMO are conditioned on the acquisition of the instantaneous channel state information (CSI), particularly at the massive BS. Using the terminology from Section~\ref{sec:CommTheory}, massive MIMO is critically dependent on the reliability/latency of the auxiliary procedures. In a mobile environment constrained by channel coherence time as well as extreme latency requirements, instantaneous CSI acquisition becomes the most severe limitation to achieve URLLC. In the general multi-device massive MIMO URLLC framework, reliability and latency are characterized by a trade-off between spatial diversity and multiplexing, as well as latency due to CSI acquisition or possibly to multiple antenna processing.

\subsubsection{Downlink: Beamforming Based on Channel Structure}

Acquisition of the instantaneous CSI at the transmitter (CSIT) is a nontrivial task for multi-antenna systems.
In FDD systems, it requires a feedback loop from the terminals inducing a significant latency.
In TDD, latency can still be reduced by exploiting channel reciprocity, but remains critical.
For URLLC it is preferable to depart from the conventional use of instantaneous CSIT, so that the question is how to benefit from the large number of transmit antennas for downlink transmission.
One solution consists of beamforming based on the multipath structure of the channel which varies on a large scale.
This structure can be estimated via the covariance matrix of the vectorial received signal from which directions of arrival or singular vectors are determined. For example, a directional beam with an angular spread encompassing a subset of the directional propagation paths can be formed.  This results in a less precise beam, sacrificing the SNR and thus the rate, but gaining in latency (short auxiliary procedure) and robustness to serve multiple terminals. Furthermore, when multiple terminals are served in the downlink, the design of a joint CSI acquisition procedure for all of them and adjusting the beams to the broadcast transmission is parallel to the ideas of joint data/metadata encoding, discussed in the previous section.

\subsubsection{Uplink: Coherent vs Non-Coherent Reception}

Since the pilots are sent in the uplink along with the data, there is less delay involved in CSIR acquisition.
Coherent multi-antenna processing can be employed at the receiver under low mobility conditions \cite{feasibility_large_arrays_urllc}.
Hence, the massive spatial multiplexing capabilities of massive MIMO can be exploited to accommodate massive connectivity in the uplink.
In URLLC, the processing delay to separate the signals from the different devices might become critical and has to be accounted for especially when the number of multiplexed devices grows.

In high-mobility scenarios, non-coherent communications may offer better performance without requiring precise CSI, even in multi-device communications.
In high mobility or low SNR scenarios, fulfilling the requirements of URLLC might require to shift to a basic TDMA system with non-coherent receivers based on energy detection (ED) \cite{noncoherent_design_performance}. In massive MIMO, ED has advantageous features, as channel and noise energy becomes deterministic and offers stable performance.

\begin{figure}[t]
    \centering
    \includegraphics[width=\linewidth]{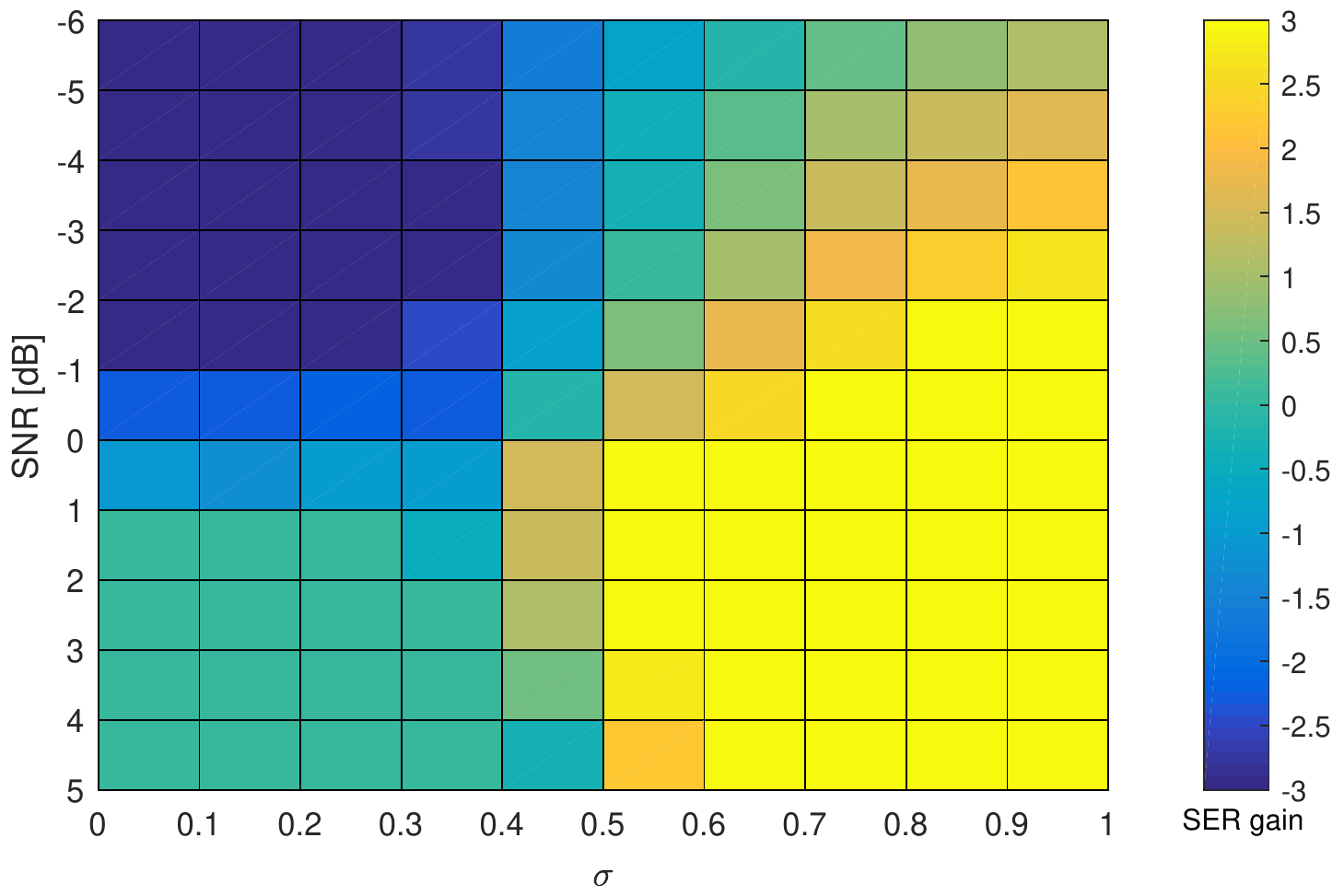}
    \caption{Heat plot showing SER gain expressed in orders of magnitude by using non-coherent ED compared to coherent MRC, with respect to SNR and the mobility index $\sigma$.}
    \label{fig:coherent_vs_noncoh}
\end{figure}

Simulations of a single-input multiple-output (SIMO) system with 128 antennas at the receiver have been performed, where the mobility is modeled as an imperfection of the received channel estimate. Fig.~\ref{fig:coherent_vs_noncoh} shows that mobility enables a couple of orders of magnitude improvement of the symbol error rate (SER) when using non-coherent ED compared to coherent maximum-ratio-combining (MRC). A mobility index $\sigma=0$ means no mobility, i.e. the channel coefficients do not change from the training symbol to the data symbol, whereas $\sigma=1$ means no correlation between the channel and the estimate.

\subsection{Interface Diversity} 

Without intervening at the physical layer, diversity can be achieved through the use of multiple links and/or communication interfaces.
Assuming that a URLLC application uses UDP, since the latency budget does not allow transport layer retransmissions, multi-interface diversity is easily achieved by duplicating the application's data packets and transmitting those through sockets attached to different communication interfaces, e.g. LTE, HSPA and Wi-Fi.
Since the experienced latency is determined by the first arriving packet, interface diversity with packet duplication (PD) leads to an increase in reliability and lowering of latency.

The concept of multi-interface diversity relates to 3GPP's dual connectivity, introduced in LTE release 12.
This technique allows for bearer or packet level split of traffic flows between a master eNB and secondary eNB for enhanced throughput.
Discussions are ongoing in 3GPP to enable data duplication for URLLC in multi-connectivity scenarios \cite{3gppTS22.261v16.1.0}.

\begin{figure}[tb]
  \centering
  \includegraphics[width=\linewidth]{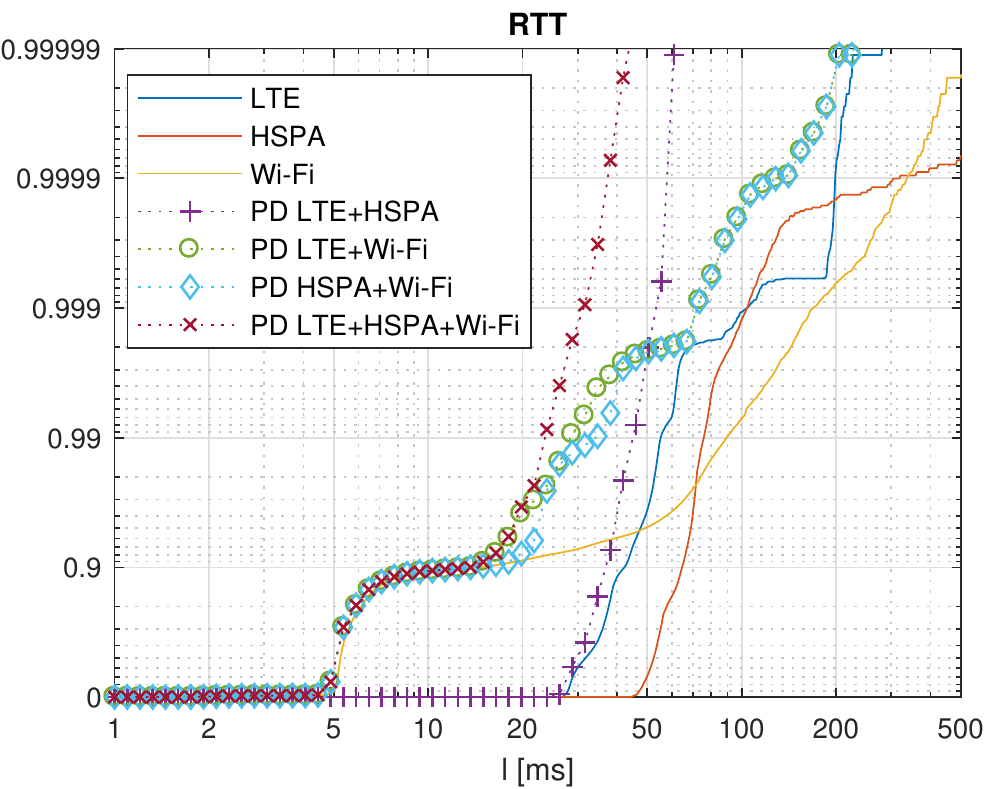}
  \caption{Achievable reliability (y-axis) for different round-trip latencies (x-axis) of packet duplication (PD) across multiple communication interfaces.}
  \label{fig:multi-interface-comm}
\end{figure}

An example of the achievable latency and reliability performance of multi-interface communication is shown in Fig. \ref{fig:multi-interface-comm}, depicting the performance of LTE, HSPA, and Wi-Fi in different single link and PD configurations.
The results are based on applying the different configurations in a simulation, where full day measurement traces of packet latency of the different technologies are played back simultaneously. The measurements have been obtained on a typical weekday at Aalborg University campus.

While the LTE+Wi-Fi and HSPA+Wi-Fi PD configurations achieve very low latency ($\leq$ 10~ms) at 0.9 reliability, both are performing relatively bad in the high reliability domains (0.9999 - 0.99999) with latencies above 100~ms.
In comparison, the LTE+HSPA and LTE+HSPA+Wi-Fi configurations achieve around 60~ms and 40~ms, respectively.

\section{Network Topology}
Another determining factor for URLLC is the way in which devices are connected, i.e. the network topology.

\subsection{Base Station Densification}
\begin{figure}[tb]
\centering
\includegraphics[width=\columnwidth]{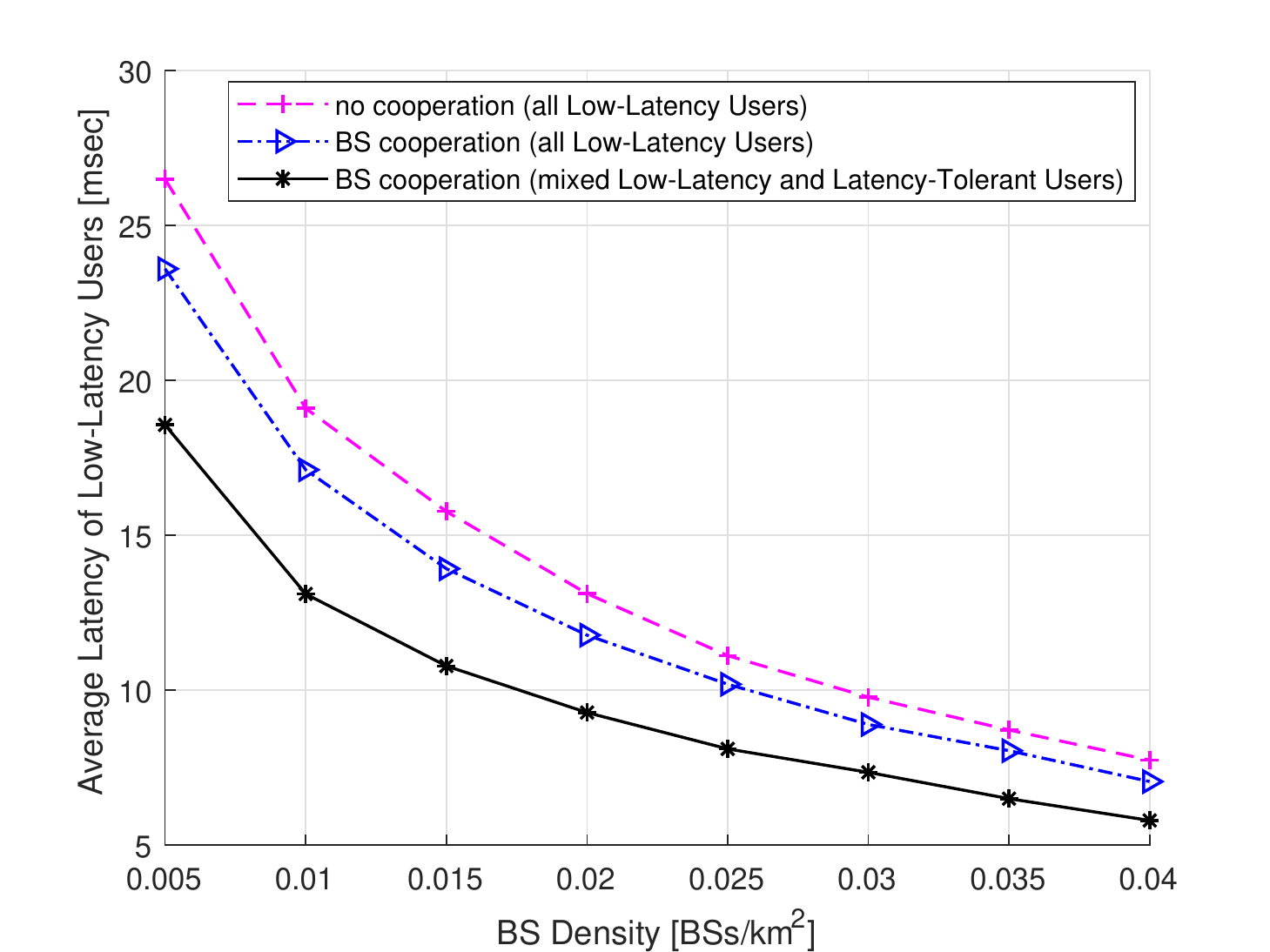}
\caption{Average latency reduction from BS densification (x-axis) and extra associations with BS cooperation (blue triangle and black star) for user density $0.01$.}
\label{F:latency_vs_bs_density}
\end{figure}

BS densification is important for achieving ubiquitous reliable connectivity, allowing users to have the best associations out of their many neighboring BSs.
This contributes to URLLC in three ways: (1) short association distance, (2) per-user resource allocation increase, and (3) multiple associations.


The decrease in BS-user association distance mitigates the propagation loss, which is important for the most severely affected users.
In the noise-limited regime, where aggregate interference is negligible compared to noise, network densification increases the desired signal power and improves the reliability.
For the interference-limited regime, the short propagation distances increase not only the desired signal power but also the interference that may be generated by numerous neighboring BSs.
Nevertheless, the desired signal power increase dominates the increase of interference due to the path-loss which follows a power-law.
Overall, network densification thereby increases signal-to-interference-plus-noise ratio (SINR) for all users \cite{UR2Cspaswin:17}.

Network densification also leads to resource reuse and increases per-user resource allocation.
This resource increment can be directly utilized for latency reduction.
Alternatively, it can be dedicated to diversity for reliability enhancement. 
Finally, network densification makes BSs more likely to have a few or even no associated users within their coverage, especially in ultra-dense network setups where the BS density exceeds user density.
Such user-void BSs are expected to be in an idle state, not sending data signals for energy-efficiency, but may provide extra associations for the URLLC users.
This, however, increases the downlink interference from the awakened BSs, which can be mitigated by cooperation between neighboring BSs.
Consider two neighboring BSs that are interconnected through a high-speed backhaul, thanks to their short inter-BS distance after densification.
In order to illustrate the concept, assume that the network features two types of users: low latency users and latency-tolerant users. 
By exchanging data signals and association information, these two BSs can serve their users concurrently without incurring interference. This can be achieved by utilizing interference cancellation or prioritizing the transmission of low-latency user \cite{dmk2017vtc}.
Fig.~\ref{F:latency_vs_bs_density} shows its effectiveness in average latency reduction.

\subsection{Device-to-Device Communication}

Traditional cellular communication follows an uplink-downlink
topology, regardless of the end-device's location. However, LTE
release~12 and 5G also supports device-to-device (D2D) communications,
where physically close devices, e.g., two vehicles, can communicate directly over a
so-called sidelink. Compared to regular uplink-downlink communication,
D2D communications benefits from a shorter link distance and fewer
hops, which is beneficial from a reliability perspective. Moreover,
since communication is direct, i.e., without intermediate nodes,
D2D has the potential to provide very low latency.

\section{Access Protocols} 
\label{sec:AccessProtocols}

Access networking represents a critical segment for development of URLLC services in cellular networks.
In this context, 3GPP follows the standard design approach exposed in Section~\ref{sec:MAPPE}, separately addressing the control-plane (i.e., metadata and auxiliary) procedures and user-plane (data) procedures, foreseeing that 5G radio access should be able to provide URLLC services with the average control-plane latency of 10~ms, the average user-plane latency of 0.5~ms, and the reliability of $99.999 \,\%$ for 32 byte long packets with latency of up to 1~ms~\cite{38.913}.
The performance of current cellular access networks is far from these goals~\cite{38.913}.
Also, some of the verticals impose reliability and latency requirements that may challenge the target 5G URLLC performance.
For instance, factory automation, an important use case in Industrie~4.0, may, according to some sources, require reliability of $1 - 10^{-9}$ (!) with 0.5~ms of user-plane end-to-end latency \cite{Ericsson}.
Note that user-plane end-to-end latency relates only to (one-way) communication delay between the source-destination pair, and, as such, is just a part of the cycle time in industrial applications; cycle time is the delay from the issuing of a command by the controller until the feedback from the actuator is received, involving all processing, actuating and sensing times~\cite{3gppTS22.261v16.1.0}.

\begin{figure}[tb]
  \centering
  \includegraphics[width=0.77\columnwidth]{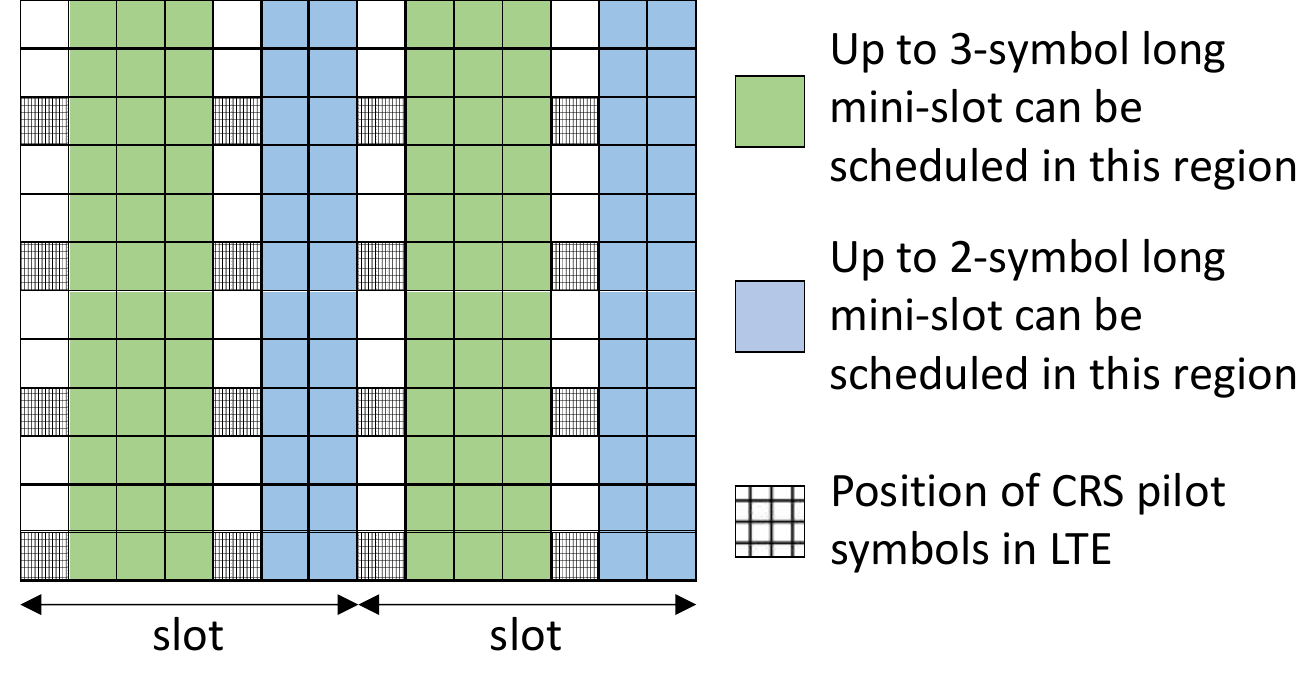}
  \caption{An example of scheduling mini-slots within a regular slot by preempting some of its content. Some restrictions should be considered, e.g. mini-slots cannot replace pilot symbols and/or control data (which is placed at the beginning of the slot).}
  \label{fig:minislot}
\end{figure}

As noted in Section~\ref{sec:LC}, the primary method to reduce latency in 5G radio access will be the use of novel numerology and shortening of transmission times.
In the downlink, latency could be further reduced by providing instant access to URLLC traffic at the expense of service performance of the other traffic types. 
Indeed, the 3GPP proposes a new unit of scheduling called mini-slot (see Fig. \ref{fig:minislot}), which can be flexibly configured to last between 1-6 OFDM symbols (while standard slots are 7 symbols long) \cite{38.802}.
Using mini-slots, the arriving URLLC data can be immediately scheduled by the BS by preempting a portion of the eMBB data operating with traditional slot-level granularity.

On the other hand, supporting URLLC requirements in the uplink is rather challenging.
Currently, uplink transmissions are subject to resource-reservation procedures with numerous stages and heavy signaling, which has a tremendous impact on latency and reliability, see Section~\ref{sec:MAPPE}.
In this respect, on-going work in 3GPP considers design of control-plane procedures exploiting pre-established contexts among devices requiring URLLC service and the BS by means of pre-configured/semi-persistent scheduling. 
However, such a solution is suitable for devices with predictable traffic patterns, but otherwise exhibits low efficiency.
Also, it applies only to the resources for the initial transmission, while any possible redundancy follows the standard, lengthy HARQ procedure.
In the following, we outline two potential approaches to access protocol redesign for traffic patterns with less determinism.

\subsection{Grant-Free Access}

The main idea of grant-free access is to skip the reservation phase.
This is a disruptive solution, 
which is by default random and non-orthogonal, involving collisions among users' transmissions.
Slotted ALOHA, the standard paradigm used for collision resolution, is not suitable for URLLC services due to its unfavorable latency/reliability trade-off.
The principal approach to achieve reliability with low latency in presence of collisions is again to rely on redundancy/diversity.
For instance, a user could devote more bandwidth and/or power to a transmission than would be necessary if there were no collisions, thus making it more robust to interference.
In such a way, multi-packet reception (MPR) can be achieved. 
Such an access scheme should also deal with collisions involving more transmissions than the MPR capability at the receiver.
In this respect, a promising approach is to proactively transmit multiple packet replicas. 
Performance in this case can be further boosted using combining techniques and/or using successive interference cancellation (SIC) on the replicas. 
This approach is also useful to combat reception errors due to noise.


Finally, a full-blown grant-free cellular uplink URLLC solution for the cases without any prior context existing between the BS and the devices should also deal with user activity detection/identification and lack of CSI.
Novel approaches advocate use of compressed sensing in this regard, where users prepend sequences to the transmitted data, which can be used both for the activity detection and channel estimation.

\subsection{Coordinated Grant-Free Access}

A certain compromise, tailored for the cases where the devices have a relatively high probability of activation, would be a protocol in which users undergo a scheduling procedure only once, followed by infrequent updates from the BS.
Scheduling information could consist of a specific access pattern, or a seed to generate it, that would tell the user in which slots to transmit the packet and its replicas, without the need for prior scheduling. The advantage of such solution is the simplified detection procedure at the BS compared to the fully random grant-free technique.
Due to the limited amount of resources and transmissions consisting of potentially several replicas per user, the access patterns need not be orthogonal.
In such case, the coordinated grant-free technique could benefit from the MPR, combining and SIC mechanisms, as described earlier.
In that case, the BS should assign access patterns in a way so that these mechanisms are best exploited to satisfy the URLLC requirements.
We conclude by noting that this approach is reminiscent of an CDMA system, where the users are assigned codes, i.e., access patterns, such that the mutual interference of the active users' transmissions is controlled. 



\section{Conclusions} 
\label{sec:conclusion}

We have formulated the communication-theoretic principles of URLLC, 
putting them into perspective of the standard models for 
communication system design that are based on the classical information/communication theory. Besides the transmission techniques applied to send data, ultra-high reliability brings in the focus the methods to send metadata and carry out the other auxiliary procedures, such as packet detection. For URLLC it is essential to invest in diversity, and the article reviewed promising approaches for doing so in the domains of device and network architecture, as well as communication protocols.
One of the key conclusions is that efficient support of URLLC requires accurate modelling and rethinking of the classical assumptions applied in communication system engineering.
The future research directions should build on detailing the design of the building blocks and combining them towards a complete URLLC solution that corresponds to a use case, such as e.g. industrial automation. For example, combining packetization/framing with the mechanism of beamforming in massive MIMO leads to the study of the tradeoff of the reliability gains from the two blocks. Integration of various diversity sources with a latency-constrained access protocol is another example of a relevant research direction implied by this paper.

\section*{Acknowledgment}
The work has partly been supported by the European Research Council (ERC Consolidator Grant nr. 648382 WILLOW), and partly by the Horizon 2020 project ONE5G (ICT-760809). 

\bibliographystyle{IEEEtran}





\end{document}